\documentclass{article}
\usepackage[dblblindworkshop,final]{neurips_2026}
\usepackage[utf8]{inputenc}
\usepackage[T1]{fontenc}
\usepackage{hyperref}
\usepackage{url}
\usepackage{booktabs}
\usepackage{amsfonts}
\usepackage{amsmath}
\usepackage{nicefrac}
\usepackage{microtype}
\usepackage{xcolor}
\usepackage{graphicx}
\usepackage{pgfplots}
\usepgfplotslibrary{groupplots}
\pgfplotsset{compat=1.18}
\title{Density-Guided Conditional Neural Processes\\for Detector Efficiency Estimation}
\newcommand{\preprintauthors}{%
  Yue Ma\textsuperscript{1}\qquad Aobo Li\textsuperscript{1,2}\\
  \textsuperscript{1}Hal{\i}c{\i}o\u{g}lu Data Science Institute\\
  \textsuperscript{2}Department of Physics\\
  University of California San Diego\\
  \texttt{y5ma@ucsd.edu}\qquad\texttt{aol002@ucsd.edu}%
}
\newcommand{\preprintpdfauthors}{Yue Ma; Aobo Li}

\workshoptitle{Machine Learning and the Physical Sciences (ML4PS 2026)}
\author{\preprintauthors}
\hypersetup{pdfauthor={\preprintpdfauthors}}
\begin{document}
\maketitle

\begin{abstract}
Searches for rare events such as neutrinoless double-beta decay
($0\nu\beta\beta$) and dark matter infer a physical rate from a handful of
events. The efficiency of the event selection is therefore critical, as it
directly impacts the sensitivity of the search.
In germanium detectors, a pulse-shape selection keeps single-site events, which
resemble the $0\nu\beta\beta$ signal, and rejects multi-site gamma-ray
backgrounds. Its efficiency is measured with calibration data and changes
sharply at spectral peaks, where the fraction of single-site events changes.
With limited calibration data, a smooth estimator erases these changes, while a
flexible one mistakes statistical fluctuations for physical structure. To
resolve this tension, we introduce a density-guided conditional neural process (CNP) that permits rapid
variation only at energies where calibration events concentrate. On public
MAJORANA calibration data, it recovers the double-escape peak, the reference
for the $0\nu\beta\beta$ signal, and it remains smooth in the continuum
around the $0\nu\beta\beta$ energy. With half as many training events, it
outperforms CNP and Attentive CNP baselines both at the peaks and in the
continuum. This reduction opens the way to continuous monitoring of the
efficiency of individual detectors in ton-scale $0\nu\beta\beta$ experiments
such as LEGEND-1000.
\end{abstract}

\section{The efficiency modeling problem}
The observation of neutrinoless double-beta decay ($0\nu\beta\beta$) would
establish that lepton number is violated and that neutrinos are Majorana
particles, with implications for the origin of the matter-antimatter
asymmetry of the universe
\citep{schechter1982neutrinoless,dolinski2019neutrinoless,agostini2023toward}.
Experiments using germanium detectors enriched in $^{76}$Ge search for this
decay as a peak at $Q_{\beta\beta}=2039$ keV and have set half-life limits
of up to $1.9\times10^{26}$ yr
\citep{agostini2020final,arnquist2023final,acharya2025first}, and the ton-scale
LEGEND-1000 experiment is in preparation \citep{abgrall2021legend}. Because
such searches infer a half-life from at most a few candidate events, the
efficiency of each analysis selection and its uncertainty enter the result
directly. The same applies to dark-matter searches
\citep{aprile2019selection} and collider measurements
\citep{cms2024tracking}.

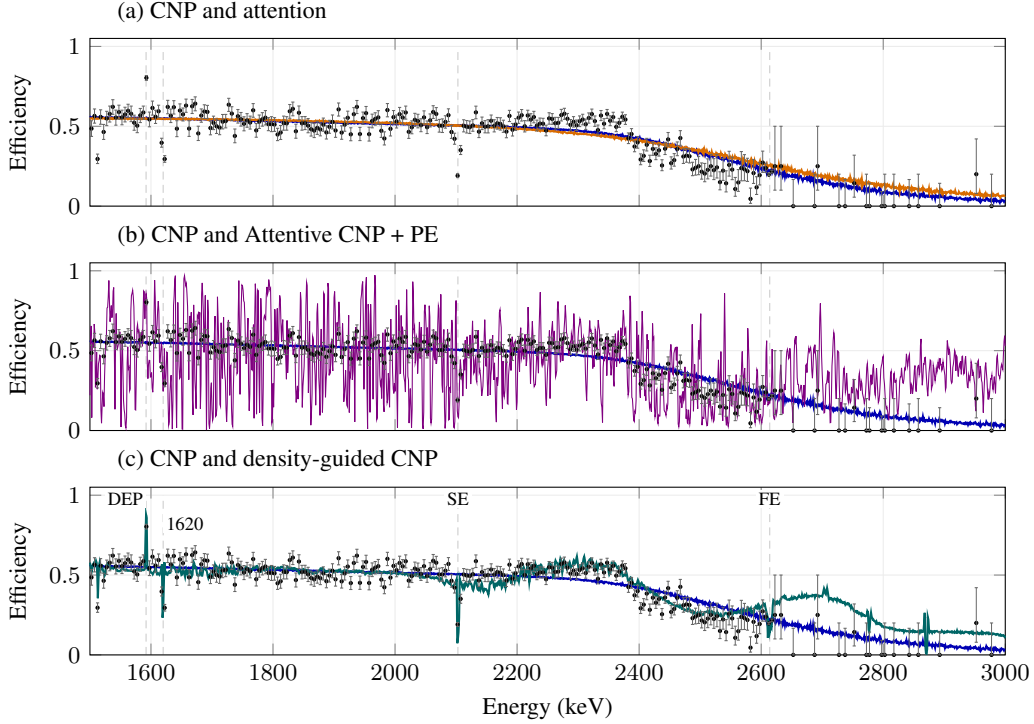
\begin{figure}[t]
\centering
\begin{tikzpicture}
\pgfplotsset{set layers=standard}
\begin{groupplot}[
 group style={group size=1 by 3,vertical sep=0.75cm},
 width=0.98\linewidth,height=3.8cm,
 xmin=1500,xmax=3000,ymin=0,ymax=1.05,ytick={0,0.5,1},
 xtick={1600,1800,2000,2200,2400,2600,2800,3000},
 scaled x ticks=false,xticklabel style={/pgf/number format/1000 sep={}},
 xlabel={},ylabel={Efficiency},
 tick label style={font=\small},label style={font=\small},
 grid=major,grid style={gray!15},
 every axis title/.style={font=\small,at={(0.02,1.04)},anchor=south west}]
\nextgroupplot[title={(a) CNP and attention},xticklabels={}]
\addplot[black,only marks,mark=*,mark size=0.6pt,on layer=axis foreground,
 error bars/y dir=both,error bars/y explicit,error bars/error bar style={black!60,line width=0.25pt}]
 table[x=energy,y=acceptance,y error plus=error_plus,y error minus=error_minus,col sep=comma]{figures/data/reference_overview.csv};
\addplot[blue!70!black,thick] table[x=energy,y=acceptance,col sep=comma]{figures/data/fit_overview_m0.csv};
\addplot[orange!85!black,thick] table[x=energy,y=acceptance,col sep=comma]{figures/data/fit_overview_m1.csv};
\begin{pgfonlayer}{axis background}
\draw[black!20,densely dashed,thin] (axis cs:1592,0)--(axis cs:1592,1);
\draw[black!20,densely dashed,thin] (axis cs:1620,0)--(axis cs:1620,1);
\draw[black!20,densely dashed,thin] (axis cs:2103,0)--(axis cs:2103,1);
\draw[black!20,densely dashed,thin] (axis cs:2614,0)--(axis cs:2614,1);
\end{pgfonlayer}
\nextgroupplot[title={(b) CNP and Attentive CNP + PE},xticklabels={}]
\addplot[black,only marks,mark=*,mark size=0.6pt,on layer=axis foreground,
 error bars/y dir=both,error bars/y explicit,error bars/error bar style={black!60,line width=0.25pt}]
 table[x=energy,y=acceptance,y error plus=error_plus,y error minus=error_minus,col sep=comma]{figures/data/reference_overview.csv};
\addplot[blue!70!black,thick] table[x=energy,y=acceptance,col sep=comma]{figures/data/fit_overview_m0.csv};
\addplot[violet] table[x=energy,y=acceptance,col sep=comma]{figures/data/fit_overview_m2.csv};
\begin{pgfonlayer}{axis background}
\draw[black!20,densely dashed,thin] (axis cs:1592,0)--(axis cs:1592,1);
\draw[black!20,densely dashed,thin] (axis cs:1620,0)--(axis cs:1620,1);
\draw[black!20,densely dashed,thin] (axis cs:2103,0)--(axis cs:2103,1);
\draw[black!20,densely dashed,thin] (axis cs:2614,0)--(axis cs:2614,1);
\end{pgfonlayer}
\nextgroupplot[title={(c) CNP and density-guided CNP},xlabel={Energy (keV)}]
\addplot[black,only marks,mark=*,mark size=0.6pt,on layer=axis foreground,
 error bars/y dir=both,error bars/y explicit,error bars/error bar style={black!60,line width=0.25pt}]
 table[x=energy,y=acceptance,y error plus=error_plus,y error minus=error_minus,col sep=comma]{figures/data/reference_overview.csv};
\addplot[blue!70!black,thick] table[x=energy,y=acceptance,col sep=comma]{figures/data/fit_overview_m0.csv};
\addplot[teal!80!black,thick] table[x=energy,y=acceptance,col sep=comma]{figures/data/fit_overview_ours.csv};
\begin{pgfonlayer}{axis background}
\draw[black!20,densely dashed,thin] (axis cs:1592,0)--(axis cs:1592,1);
\draw[black!20,densely dashed,thin] (axis cs:1620,0)--(axis cs:1620,1);
\draw[black!20,densely dashed,thin] (axis cs:2103,0)--(axis cs:2103,1);
\draw[black!20,densely dashed,thin] (axis cs:2614,0)--(axis cs:2614,1);
\end{pgfonlayer}
\node[font=\scriptsize,anchor=south east,fill=white,inner sep=1pt] at (axis cs:1592,0.92) {DEP};
\node[font=\scriptsize,anchor=north west,fill=white,inner sep=1pt] at (axis cs:1620,0.88) {1620};
\node[font=\scriptsize,anchor=south,fill=white,inner sep=1pt] at (axis cs:2103,0.92) {SEP};
\node[font=\scriptsize,anchor=south,fill=white,inner sep=1pt] at (axis cs:2614,0.92) {FEP};
\end{groupplot}
\end{tikzpicture}
\caption{Models use 5k training and 500 context events. Black reference points have
5-keV bins ($n\geq4$) and approximate 68\% Wilson intervals. Blue, orange, purple, and teal
denote CNP, Attentive CNP, Attentive CNP + PE, and ours.}
\label{fig:fits}
\end{figure}

A $0\nu\beta\beta$ decay deposits its energy within about a millimeter, whereas
gamma-ray backgrounds typically scatter and deposit energy at several sites.
Germanium experiments exploit this difference through pulse-shape
discrimination, which retains single-site events and rejects multi-site events
\citep{budjas2009pulse,alvis2019multisite,agostini2022pulse}. The efficiency of
this selection is determined from $^{228}$Th calibration data
\citep{abgrall2017calibration,arnquist2023data}, whose spectrum contains event
populations of known topology. Events in the double-escape peak (DEP) at 1592
keV are single-site and serve as the reference for the signal, whereas the
single-escape peak (SEP) at 2103 keV consists mostly of multi-site events and
the region around $Q_{\beta\beta}$ is a continuum of Compton-scattered events.
The efficiency consequently changes sharply at these peaks and at the
full-energy peaks (FEP) at 2614 keV from $^{208}$Tl and 1620 keV from
$^{212}$Bi, while it varies slowly across the continuum
(Figure~\ref{fig:fits}).

The precision of this measurement is limited by calibration statistics.
MAJORANA required about 12 hours of calibration to collect sufficient DEP
events for each detector, and its hour-long calibrations were insufficient for
this purpose \citep{alvis2019multisite}. Estimating the efficiency from fewer
events requires a model that follows the peaks without introducing spurious
structure in the continuum, which would bias the efficiency in the signal
region. Machine learning is already used to classify pulse shapes in germanium
detectors \citep{holl2019deep,arnquist2023interpretable}, and conditional
neural processes (CNPs) have served as data-efficient surrogates in detector
design \citep{schuetz2025resum}. A CNP predicts a function from a small set of
labeled examples, called the context \citep{garnelo2018cnp,garnelo2018neural}.
Standard and attentive CNPs \citep{kim2019anp,vaswani2017attention} are
nevertheless too smooth to follow the peaks, since neural networks learn slowly
varying functions first \citep{rahaman2019spectral}. Positional encoding (PE)
removes this limitation \citep{tancik2020fourier} but permits rapid variation
at all energies, so that the model also fits statistical fluctuations in the
continuum, whereas adaptive encodings such as SAPE leave the choice of location
to the optimization \citep{hertz2021sape}.

We let the physics decide instead. The peaks are where calibration events
concentrate in energy, so we use the event density to permit rapid variation
only there. The resulting model recovers the DEP while remaining smooth around
$Q_{\beta\beta}$ (Table~\ref{tab:agreement}). It outperforms CNP and Attentive
CNP trained on twice as many events (Table~\ref{tab:budget}), and an ablation
attributes the gain to the density-dependent gate (Table~\ref{tab:mechanism}).

\section{Density-guided efficiency estimation}
The data release labels each event by whether it passes a standard selection
that rejects multi-site energy deposits \citep{arnquist2023data}. A
convolutional neural network (CNN) is trained on the detector waveforms to
reproduce this label. We freeze the network, take its output as a score $s$,
and estimate the efficiency $\beta(E;T)=\Pr(s\geq T\mid E)$ of the cut
$s\geq T$.
The CNP receives context events with known outcomes
$\mathcal C_T=\{((e_i,T),X_i)\}_{i=1}^{n_C}$, where
$X_i=\mathbf1[s_i\geq T]$ and $e=(E-E_{\min})/\Delta E$.
The energy origin $E_{\min}$ is 500 keV and scale $\Delta E$ is 2500 keV.
Training varies the cut and context, while evaluation uses one fixed cut.

\paragraph{Physics-informed flexibility.}
The flexibility of the model is physics-informed in the sense that rapid
variation is permitted only where the underlying physics produces narrow
structure. In a calibration spectrum, such structure appears as a local
concentration of events in energy. We quantify this concentration by comparing
the event density at a narrow scale with that at a broad scale. For the energies
$E_p$ of the training events $\mathcal P$, we define the density ratio
\begin{equation}
 A_\sigma(E)=\sum_{p\in\mathcal P}e^{-(E-E_p)^2/(2\sigma^2)},\qquad
 R(E)=\frac{\sigma_g A_{\sigma_l}(E)}
 {\sigma_l[A_{\sigma_g}(E)+\varepsilon_d]}.
 \label{eq:density}
\end{equation}
We set $\sigma_l=1$ keV near the Gaussian detector-resolution scale
\citep{arnquist2023final}. The broader $\sigma_g=50$ keV separates narrow
concentrations from broader Compton structure, but can blend nearby lines.
The stabilizer $\varepsilon_d=10^{-5}$ prevents division by zero.
$R$ is near one in a populated flat spectrum and rises at narrow peaks.
Fourier features $\gamma_\ell(e)=[\sin(2^\ell\pi e),\cos(2^\ell\pi e)]$
for $\ell=0,\ldots,L-1$ span ungated periods of about 10 to 5,000 keV.
We set their cutoff and weights using
\begin{equation}
 \lambda(E)=\kappa+(L-\kappa)\operatorname{sigmoid}(\alpha_R[R(E)-R_0]),\qquad
 w_\ell(E)=\operatorname{sigmoid}(\alpha_F[\lambda(E)-\ell]).
 \label{eq:gate}
\end{equation}
Inspecting the spectrum and density ratio motivated $R_0=3$, above
continuum fluctuations and below prominent peak contrasts. The learned floor $\kappa$ lies between one and
five, starts at three, and controls background flexibility.
We use $L=10$ levels. Fixed slopes $\alpha_R=10$ and $\alpha_F=5$ control the
contrast transition and frequency taper.
The gated features $w_\ell\gamma_\ell$ enter the decoder, the network that
produces the efficiency estimate. The decoder is a multilayer perceptron with
three hidden layers. Its input is $[r,e,T,\{w_\ell\gamma_\ell\}_\ell,R]$, where
$r$ is the context summary defined in the next paragraph. A sigmoid maps the
decoder output to $\beta$. The density is always computed from the
matching training pool and never from reference events.

\paragraph{Density-guided context aggregation.}
An encoder network maps each context event to a vector $v_i$ computed from its
ungated Fourier features $\gamma_\ell$, its cut, and its outcome. The gate of
Eq.~\ref{eq:gate} therefore acts only on the decoder input. The context summary
for a query at energy $E$ is $r(E)=W_o\sum_i\omega_i(E)\,v_i$, where $W_o$ is a
learned projection and the attention weights are
\begin{equation}
 \omega_i(E)=\operatorname{softmax}_i\!\left[
 \frac{q^\top k_i}{\sqrt{d_{\mathrm{att}}}\,\tau(E)}
 -\frac{(E-E_i)^2}{2h(E)^2}\right].
 \label{eq:attention}
\end{equation}
The query $q$ and the keys $k_i$ are linear projections of the raw coordinates
$(e,T)$ with dimension $d_{\mathrm{att}}=128$. The width $h(E)$ limits the
energy range over which context events are pooled, and the temperature
$\tau(E)$ determines how selectively the model weights the events within that
range. Two small networks compute $h$ and $\tau$ from $\log A_{\sigma_l}(E)$
and $\log A_{\sigma_g}(E)$ of Eq.~\ref{eq:density}. Their outputs are bounded
to the ranges listed in Appendix~\ref{app:math}.

\begin{table}[t]
\centering
\caption{\textbf{Local recovery and continuum agreement.}
Entries give $C_2$ (\%), the percentage of bins in which the prediction agrees
with the reference within twice its statistical uncertainty
(Section~\ref{sec:results}). Higher values are better. Peaks averages the four
peak cores, Continuum averages two windows between the peaks, and Overall
covers 500 to 3000 keV. The joint columns show the mean and, in parentheses,
the standard deviation over seeds. Bold marks the highest value in each column.
Appendix~\ref{app:evidence} defines the bins.}
\label{tab:agreement}
\begin{tabular}{@{}lrrrrrrr@{}}
\toprule
 & \multicolumn{2}{c}{Joint summary} & \multicolumn{4}{c}{Individual peak cores (keV)} & \\
\cmidrule(lr){2-3}\cmidrule(lr){4-7}
 & & & $^{208}$Tl & $^{212}$Bi & $^{208}$Tl & $^{208}$Tl & \\
 & & & DEP & FEP & SEP & FEP & \\
Model & Peaks & Continuum & 1592 & 1620 & 2103 & 2614 & Overall\\
\midrule
CNP & 23 (9) & 73 (7) & 50 & 0 & 0 & 43 & \textbf{80} \\
Attentive CNP & 18 (7) & 66 (9) & 50 & 0 & 0 & 20 & 75 \\
Attentive CNP + PE & 41 (3) & 37 (6) & 17 & \textbf{60} & \textbf{72} & 17 & 36 \\
\textbf{Ours} & \textbf{45} (13) & \textbf{89} (1) & \textbf{100} & 50 & 17 & 13 & \textbf{80} \\
\addlinespace[2pt]
Kernel & 12 (10) & 63 (20) & 35 & 0 & 5 & 10 & 57 \\
Bernoulli GP & 11 (11) & 61 (19) & 35 & 5 & 5 & 0 & 55 \\
\addlinespace[2pt]
Kernel (pooled) & 36 (4) & 87 (2) & 30 & 20 & 45 & \textbf{50} & 75 \\
\bottomrule

\end{tabular}
\end{table}

\section{Data budgets and evaluation}
The CNN is trained for 50 epochs on 18.9k events, which are 5.0\% of a pool of
$3.8\times10^5$ events. The efficiency model is trained on the energies and
scores of 2k, 5k, or 10k of these events, which are 0.53\%, 1.3\%, and 2.7\%
of the pool, and the density ratio is computed from the same subset. The cut $T\approx0.54$ and the model settings are fixed on a
separate set of 7,074 events. At evaluation, each model receives 500 context events
drawn from a pool of 20k events, and its predictions are compared with a
reference of 114,400 events. The training, context, and reference events are
disjoint.

All neural models use identical network widths, dropout, learning rate, and
training schedule, with 3,000 Adam steps and batches of 16 tasks sampled with
replacement across eligible energy bins. They differ only in their
architectural components, and no model was tuned individually. The constants
specific to our model, such as $R_0$, were fixed by full-pool exploration and
development Brier selection before the budget comparison.

Two classical baselines are fitted to the same 500 context events without any
training events. The first is Gaussian kernel regression
\citep{nadaraya1964regression}, and the second is a Gaussian process (GP)
with a Bernoulli likelihood \citep{rasmussen2006gpml}. Their
bandwidth and kernel are selected on development data. A third baseline, the
pooled kernel, applies the kernel regression to the context events together
with the classifier's 18.9k training events.

Each configuration is run with three random seeds and ten context sets, and
every prediction averages 50 dropout passes. The 5k budget is repeated on two
further training subsets. The tables report means over these runs, and
Figure~\ref{fig:fits} shows the mean prediction for the first 5k subset. Appendices~\ref{app:math}
and~\ref{app:evidence} give the architecture, the data allocation, and the
remaining settings.

\section{Reconstruction quality and data efficiency}
\label{sec:results}

\paragraph{The efficiency follows the event topology.}
The measured efficiency is about 0.52 in the continuum. It rises to 0.80 at the
DEP, where events are single-site, and falls to 0.19 at the SEP, where they are
mostly multi-site (Figure~\ref{fig:fits}). CNP and Attentive CNP miss both
features and predict about 0.55 at the DEP and 0.50 at the SEP. Attentive CNP + PE
reaches 0.72 and 0.30, but its prediction fluctuates throughout the continuum,
including around $Q_{\beta\beta}$. Our model reaches 0.71 and 0.25 and remains
smooth in the continuum.

\paragraph{Agreement with the reference.}
We quantify these observations in 5-keV bins. $C_2$ is the percentage of bins
in which the prediction differs from the measured efficiency by at most twice
its statistical uncertainty, taken as the half-width of a 68\% Wilson interval
\citep{wilson1927inference}. We count bins instead of summing a $\chi^2$
because the predictions carry no calibrated uncertainty and because a sum would
be dominated by the few peak bins with the highest counts. In the continuum on
both sides of $Q_{\beta\beta}$, our model scores 89\%, compared with 37\% for
Attentive CNP + PE (Table~\ref{tab:agreement}). At the DEP it scores 100\%. Its
scores at the SEP and the FEPs are lower, and Attentive CNP + PE scores higher
there. These peaks have the highest counts, which reduces the tolerance to
about 0.01 at 2614 keV, compared with 0.07 to 0.09 in the continuum. The two
classical baselines score lower in both summaries. 

\paragraph{Training-data efficiency.}
With 5k training events, our model exceeds CNP and Attentive CNP trained on 10k
events in both summaries and on all three subsets (Table~\ref{tab:budget}). It
therefore uses \emph{half the training events}. Its margins over the 10k models in (Peaks, Continuum) are
$(+23\%, +12\%)$ against CNP and $(+19\%, +15\%)$ against Attentive CNP.
Against 10k PE the margins are $(-6\%, +40\%)$, so it trades peak agreement for
continuum recovery. With 2k events our model does not reach these scores. Between 5k and 10k its
scores change by less than the seed SD without retuning.

\begin{table}[t]
\centering
\caption{\textbf{Reconstruction with fewer training events.}
Entries give mean $C_2$ (\%) with seed SD in parentheses, using the regions
in Table~\ref{tab:agreement}. The classifier and 500-event contexts are fixed.}
\label{tab:budget}
\begin{tabular}{@{}lrrrrrr@{}}
\toprule
 & \multicolumn{2}{c}{2,000} & \multicolumn{2}{c}{5,000} & \multicolumn{2}{c}{10,000}\\
\cmidrule(lr){2-3}\cmidrule(lr){4-5}\cmidrule(lr){6-7}
Model & Peaks & Continuum & Peaks & Continuum & Peaks & Continuum\\
\midrule
CNP & \textbf{27} (1) & \textbf{82} (9) & 23 (9) & 73 (7) & 22 (8) & 77 (3) \\
Attentive CNP & 12 (0) & 78 (8) & 18 (7) & 66 (9) & 26 (13) & 74 (3) \\
Attentive CNP + PE & 22 (11) & 26 (2) & 41 (3) & 37 (6) & \textbf{51} (5) & 49 (7) \\
\textbf{Ours} & 13 (9) & 42 (2) & \textbf{45} (13) & \textbf{89} (1) & 42 (14) & \textbf{88} (2) \\
\bottomrule

\end{tabular}
\end{table}

\paragraph{Which density pathway matters?}
Table~\ref{tab:mechanism} separates the two density-dependent components. One
control uses a single learned frequency cutoff, and another uses constant
attention width and temperature. The density-free control applies both changes
and removes $R$. Relative to a global cutoff, the adaptive gate gains $24.2\%$
at peaks and $7.1\%$ in the continuum, positive in all three seed means,
whereas constant attention changes the scores by less than $1\%$. The gain therefore arises from the adaptive gate
rather than from adapting attention (Appendix~\ref{app:math}).

\begin{table}[ht]
\centering
\caption{\textbf{Adaptive decoder gating drives the gains.} Variants share the 5k subset,
seeds, and 500-event contexts. Entries are mean $C_2$ (\%) with seed SD.
Bold marks the highest value in each column.}
\label{tab:mechanism}
\begin{tabular}{@{}lrr@{}}
\toprule
Variant & Peaks & Continuum\\
\midrule
Full model (ours) & 45 (13) & \textbf{89} (1) \\
Global decoder gate & 21 (7) & 82 (3) \\
Global attention & \textbf{46} (16) & \textbf{89} (1) \\
Global gate + attention & 23 (3) & 82 (3) \\
Density-free global & 20 (1) & 87 (4) \\
\bottomrule

\end{tabular}
\end{table}

\section{Discussion}
The efficiency of a multi-site rejection changes sharply only where the event
topology changes, and the calibration spectrum shows where that is. With this
knowledge, our model recovers the DEP, the reference for the $0\nu\beta\beta$
signal, without inventing structure in the continuum around $Q_{\beta\beta}$,
and it outperforms CNP baselines trained on twice as many events.

These results bear on the calibration budget of rare-event searches. MAJORANA
needed about 12 hours of calibration to tune this selection for each detector,
whereas its shorter and more frequent calibrations could only monitor stability
\citep{alvis2019multisite,abgrall2017calibration}. An estimator that requires
fewer events could extract the efficiency from these short calibrations, which
would allow the efficiency of each detector to be monitored continuously rather
than determined once per data set.
Within the germanium program, this applies directly to LEGEND-200
\citep{acharya2025first} and LEGEND-1000 \citep{abgrall2021legend}, which rely
on the same kind of pulse-shape discrimination
\citep{budjas2009pulse,alvis2019multisite,agostini2022pulse}.

The method is not tied to germanium or to pulse shapes. It requires only that
the efficiency change sharply at locations that can be identified in advance
and vary slowly elsewhere. This structure is common in the next generation of
rare-event searches, including $0\nu\beta\beta$ experiments with other isotopes
such as nEXO \citep{adhikari2022nexo}, and
dark-matter experiments such as XLZD \citep{aalbers2025xlzd} and DarkSide-20k
\citep{aalseth2018darkside}, where selection efficiencies are likewise
calibrated with dedicated sources and change rapidly near energy thresholds.
Local features of this kind arise wherever the detector response changes
abruptly with a physical variable. Examples include energy thresholds, escape
and absorption lines, the boundaries of the active volume, surface layers in
which charge collection is incomplete \citep{arnquist2023data}, and changes of
operating conditions over time. In each case the location of the feature
follows from the physics of the detector and can replace the event density as
the guide. The approach therefore applies to any measurement in which physics
determines where the response may change abruptly, and it offers a general way
to build this knowledge into data-efficient models of detector response.

\begin{ack}
This work is supported by the U.S. Department of Energy under award number
DE-AC02-05CH11231.
\end{ack}

\bibliographystyle{plainnat}
\bibliography{references}

\begin{thebibliography}{30}
\providecommand{\natexlab}[1]{#1}
\providecommand{\url}[1]{\texttt{#1}}
\expandafter\ifx\csname urlstyle\endcsname\relax
  \providecommand{\doi}[1]{doi: #1}\else
  \providecommand{\doi}{doi: \begingroup \urlstyle{rm}\Url}\fi

\bibitem[Aalbers et~al.(2025)]{aalbers2025xlzd}
J.~Aalbers et~al.
\newblock The {XLZD} design book: Towards the next-generation liquid xenon
  observatory for dark matter and neutrino physics.
\newblock \emph{The European Physical Journal C}, 85:\penalty0 1192, 2025.
\newblock \doi{10.1140/epjc/s10052-025-14810-w}.

\bibitem[Aalseth et~al.(2018)]{aalseth2018darkside}
C.~E. Aalseth et~al.
\newblock {DarkSide-20k}: A 20 tonne two-phase {LAr} {TPC} for direct dark
  matter detection at {LNGS}.
\newblock \emph{The European Physical Journal Plus}, 133:\penalty0 131, 2018.
\newblock \doi{10.1140/epjp/i2018-11973-4}.

\bibitem[Abgrall et~al.(2017)]{abgrall2017calibration}
N.~Abgrall et~al.
\newblock The {Majorana Demonstrator} calibration system.
\newblock \emph{Nuclear Instruments and Methods in Physics Research Section A},
  872:\penalty0 16--22, 2017.
\newblock \doi{10.1016/j.nima.2017.08.005}.

\bibitem[Abgrall et~al.(2021)]{abgrall2021legend}
N.~Abgrall et~al.
\newblock {LEGEND-1000} preconceptual design report, 2021.
\newblock URL \url{https://arxiv.org/abs/2107.11462}.

\bibitem[Acharya et~al.(2026)]{acharya2025first}
H.~Acharya et~al.
\newblock First results on the search for lepton number violating neutrinoless
  double-{$\beta$} decay with the {LEGEND-200} experiment.
\newblock \emph{Physical Review Letters}, 136:\penalty0 022701, 2026.
\newblock \doi{10.1103/25tk-nctn}.
\newblock arXiv:2505.10440.

\bibitem[Adhikari et~al.(2022)]{adhikari2022nexo}
G.~Adhikari et~al.
\newblock {nEXO}: Neutrinoless double beta decay search beyond $10^{28}$ year
  half-life sensitivity.
\newblock \emph{Journal of Physics G: Nuclear and Particle Physics},
  49:\penalty0 015104, 2022.
\newblock \doi{10.1088/1361-6471/ac3631}.

\bibitem[Agostini et~al.(2020)]{agostini2020final}
M.~Agostini et~al.
\newblock Final results of {GERDA} on the search for neutrinoless
  double-{$\beta$} decay.
\newblock \emph{Physical Review Letters}, 125:\penalty0 252502, 2020.
\newblock \doi{10.1103/PhysRevLett.125.252502}.

\bibitem[Agostini et~al.(2022)]{agostini2022pulse}
M.~Agostini et~al.
\newblock Pulse shape analysis in {GERDA} {Phase II}.
\newblock \emph{The European Physical Journal C}, 82:\penalty0 284, 2022.
\newblock \doi{10.1140/epjc/s10052-022-10163-w}.

\bibitem[Agostini et~al.(2023)Agostini, Benato, Detwiler, Men{\'e}ndez, and
  Vissani]{agostini2023toward}
Matteo Agostini, Giovanni Benato, Jason~A. Detwiler, Javier Men{\'e}ndez, and
  Francesco Vissani.
\newblock Toward the discovery of matter creation with neutrinoless
  {$\beta\beta$} decay.
\newblock \emph{Reviews of Modern Physics}, 95:\penalty0 025002, 2023.
\newblock \doi{10.1103/RevModPhys.95.025002}.

\bibitem[Alvis et~al.(2019)]{alvis2019multisite}
S.~I. Alvis et~al.
\newblock Multisite event discrimination for the {Majorana Demonstrator}.
\newblock \emph{Physical Review C}, 99:\penalty0 065501, 2019.
\newblock \doi{10.1103/PhysRevC.99.065501}.

\bibitem[Aprile et~al.(2019)]{aprile2019selection}
E.~Aprile et~al.
\newblock {XENON1T} dark matter data analysis: Signal reconstruction,
  calibration, and event selection.
\newblock \emph{Physical Review D}, 100:\penalty0 052014, 2019.
\newblock \doi{10.1103/PhysRevD.100.052014}.

\bibitem[Arnquist et~al.(2023{\natexlab{a}})]{arnquist2023data}
I.~J. Arnquist et~al.
\newblock {Majorana Demonstrator} data release for {AI/ML} applications,
  2023{\natexlab{a}}.
\newblock URL \url{https://arxiv.org/abs/2308.10856}.

\bibitem[Arnquist et~al.(2023{\natexlab{b}})]{arnquist2023final}
I.~J. Arnquist et~al.
\newblock Final result of the {Majorana Demonstrator}'s search for neutrinoless
  double-{$\beta$} decay in {$^{76}$Ge}.
\newblock \emph{Physical Review Letters}, 130:\penalty0 062501,
  2023{\natexlab{b}}.
\newblock \doi{10.1103/PhysRevLett.130.062501}.

\bibitem[Arnquist et~al.(2023{\natexlab{c}})]{arnquist2023interpretable}
I.~J. Arnquist et~al.
\newblock Interpretable boosted-decision-tree analysis for the {Majorana
  Demonstrator}.
\newblock \emph{Physical Review C}, 107:\penalty0 014321, 2023{\natexlab{c}}.
\newblock \doi{10.1103/PhysRevC.107.014321}.

\bibitem[Budj{\'a}{\v{s}} et~al.(2009)Budj{\'a}{\v{s}}, Heider, Chkvorets,
  Khanbekov, and Sch{\"o}nert]{budjas2009pulse}
Du{\v{s}}an Budj{\'a}{\v{s}}, Marik~Barnab{\'e} Heider, Oleg Chkvorets, Nikita
  Khanbekov, and Stefan Sch{\"o}nert.
\newblock Pulse shape discrimination studies with a broad-energy germanium
  detector for signal identification and background suppression in the {GERDA}
  double beta decay experiment.
\newblock \emph{Journal of Instrumentation}, 4:\penalty0 P10007, 2009.
\newblock \doi{10.1088/1748-0221/4/10/P10007}.

\bibitem[{CMS Collaboration}(2024)]{cms2024tracking}
{CMS Collaboration}.
\newblock Tracking performance using tag and probe with {$Z \rightarrow
  \mu^{+}\mu^{-}$} in 2022 and 2023.
\newblock CMS Detector Performance Summary CMS-DP-2024-054, CERN, 2024.
\newblock URL \url{https://cds.cern.ch/record/2904367/files/DP2024_054.pdf}.

\bibitem[Dolinski et~al.(2019)Dolinski, Poon, and
  Rodejohann]{dolinski2019neutrinoless}
Michelle~J. Dolinski, Alan W.~P. Poon, and Werner Rodejohann.
\newblock Neutrinoless double-beta decay: Status and prospects.
\newblock \emph{Annual Review of Nuclear and Particle Science}, 69:\penalty0
  219--251, 2019.
\newblock \doi{10.1146/annurev-nucl-101918-023407}.

\bibitem[Garnelo et~al.(2018{\natexlab{a}})Garnelo, Rosenbaum, Maddison,
  Ramalho, Saxton, Shanahan, Teh, Rezende, and Eslami]{garnelo2018cnp}
Marta Garnelo, Dan Rosenbaum, Christopher Maddison, Tiago Ramalho, David
  Saxton, Murray Shanahan, Yee~Whye Teh, Danilo Rezende, and S.~M.~Ali Eslami.
\newblock Conditional neural processes.
\newblock In \emph{Proceedings of the 35th International Conference on Machine
  Learning}, volume~80 of \emph{Proceedings of Machine Learning Research},
  pages 1704--1713. PMLR, 2018{\natexlab{a}}.
\newblock URL \url{https://proceedings.mlr.press/v80/garnelo18a.html}.

\bibitem[Garnelo et~al.(2018{\natexlab{b}})Garnelo, Schwarz, Rosenbaum, Viola,
  Rezende, Eslami, and Teh]{garnelo2018neural}
Marta Garnelo, Jonathan Schwarz, Dan Rosenbaum, Fabio Viola, Danilo~J. Rezende,
  S.~M.~Ali Eslami, and Yee~Whye Teh.
\newblock Neural processes, 2018{\natexlab{b}}.
\newblock URL \url{https://arxiv.org/abs/1807.01622}.

\bibitem[Hertz et~al.(2021)Hertz, Perel, Giryes, Sorkine-Hornung, and
  Cohen-Or]{hertz2021sape}
Amir Hertz, Or~Perel, Raja Giryes, Olga Sorkine-Hornung, and Daniel Cohen-Or.
\newblock {SAPE}: Spatially-adaptive progressive encoding for neural
  optimization.
\newblock In \emph{Advances in Neural Information Processing Systems},
  volume~34, pages 8820--8832, 2021.
\newblock URL
  \url{https://proceedings.neurips.cc/paper/2021/hash/4a06d868d044c50af0cf9bc82d2fc19f-Abstract.html}.

\bibitem[Holl et~al.(2019)Holl, Hauertmann, Majorovits, Schulz, Schuster, and
  Zsigmond]{holl2019deep}
P.~Holl, L.~Hauertmann, B.~Majorovits, O.~Schulz, M.~Schuster, and A.~J.
  Zsigmond.
\newblock Deep learning based pulse shape discrimination for germanium
  detectors.
\newblock \emph{The European Physical Journal C}, 79:\penalty0 450, 2019.
\newblock \doi{10.1140/epjc/s10052-019-6869-2}.

\bibitem[Kim et~al.(2019)Kim, Mnih, Schwarz, Garnelo, Eslami, Rosenbaum,
  Vinyals, and Teh]{kim2019anp}
Hyunjik Kim, Andriy Mnih, Jonathan Schwarz, Marta Garnelo, Ali Eslami, Dan
  Rosenbaum, Oriol Vinyals, and Yee~Whye Teh.
\newblock Attentive neural processes.
\newblock In \emph{International Conference on Learning Representations}, 2019.
\newblock URL \url{https://arxiv.org/abs/1901.05761}.

\bibitem[Nadaraya(1964)]{nadaraya1964regression}
E.~A. Nadaraya.
\newblock On estimating regression.
\newblock \emph{Theory of Probability \& Its Applications}, 9\penalty0
  (1):\penalty0 141--142, 1964.
\newblock \doi{10.1137/1109020}.

\bibitem[Rahaman et~al.(2019)Rahaman, Baratin, Arpit, Draxler, Lin, Hamprecht,
  Bengio, and Courville]{rahaman2019spectral}
Nasim Rahaman, Aristide Baratin, Devansh Arpit, Felix Draxler, Min Lin, Fred
  Hamprecht, Yoshua Bengio, and Aaron Courville.
\newblock On the spectral bias of neural networks.
\newblock In \emph{Proceedings of the 36th International Conference on Machine
  Learning}, volume~97 of \emph{Proceedings of Machine Learning Research},
  pages 5301--5310. PMLR, 2019.
\newblock URL \url{https://proceedings.mlr.press/v97/rahaman19a.html}.

\bibitem[Rasmussen and Williams(2006)]{rasmussen2006gpml}
Carl~Edward Rasmussen and Christopher K.~I. Williams.
\newblock \emph{Gaussian Processes for Machine Learning}.
\newblock The MIT Press, 2006.
\newblock ISBN 0-262-18253-X.
\newblock URL \url{https://gaussianprocess.org/gpml/}.

\bibitem[Schechter and Valle(1982)]{schechter1982neutrinoless}
J.~Schechter and J.~W.~F. Valle.
\newblock Neutrinoless double-{$\beta$} decay in {SU(2)$\times$U(1)} theories.
\newblock \emph{Physical Review D}, 25:\penalty0 2951--2954, 1982.
\newblock \doi{10.1103/PhysRevD.25.2951}.

\bibitem[Schuetz et~al.(2025)Schuetz, Poon, and Li]{schuetz2025resum}
Ann-Kathrin Schuetz, Alan Poon, and Aobo Li.
\newblock {RESuM}: A rare event surrogate model for physics detector design.
\newblock In \emph{International Conference on Learning Representations}, 2025.
\newblock URL
  \url{https://proceedings.iclr.cc/paper_files/paper/2025/hash/c3909e3abe8ebdb20c42a42ce0bc907d-Abstract-Conference.html}.

\bibitem[Tancik et~al.(2020)Tancik, Srinivasan, Mildenhall, Fridovich-Keil,
  Raghavan, Singhal, Ramamoorthi, Barron, and Ng]{tancik2020fourier}
Matthew Tancik, Pratul Srinivasan, Ben Mildenhall, Sara Fridovich-Keil, Nithin
  Raghavan, Utkarsh Singhal, Ravi Ramamoorthi, Jonathan Barron, and Ren Ng.
\newblock Fourier features let networks learn high frequency functions in low
  dimensional domains.
\newblock In \emph{Advances in Neural Information Processing Systems},
  volume~33, pages 7537--7547, 2020.
\newblock URL
  \url{https://proceedings.neurips.cc/paper_files/paper/2020/hash/55053683268957697aa39fba6f231c68-Abstract.html}.

\bibitem[Vaswani et~al.(2017)Vaswani, Shazeer, Parmar, Uszkoreit, Jones, Gomez,
  Kaiser, and Polosukhin]{vaswani2017attention}
Ashish Vaswani, Noam Shazeer, Niki Parmar, Jakob Uszkoreit, Llion Jones,
  Aidan~N. Gomez, {\L}ukasz Kaiser, and Illia Polosukhin.
\newblock Attention is all you need.
\newblock In \emph{Advances in Neural Information Processing Systems},
  volume~30, 2017.
\newblock URL \url{https://arxiv.org/abs/1706.03762}.

\bibitem[Wilson(1927)]{wilson1927inference}
Edwin~B. Wilson.
\newblock Probable inference, the law of succession, and statistical inference.
\newblock \emph{Journal of the American Statistical Association}, 22\penalty0
  (158):\penalty0 209--212, 1927.
\newblock \doi{10.1080/01621459.1927.10502953}.

\end{thebibliography}
\clearpage
\appendix
\numberwithin{equation}{section}
\section{Model architecture and training}
\label{app:math}

\subsection{Inputs and context representation}
The model estimates efficiency conditional on a fixed classifier and a threshold
$T$. An event has normalized energy $e=(E-E_{\min})/\Delta E$ and a binary outcome
$X=\mathbf{1}[s\geq T]$. The context encoder combines energy features, the threshold,
and the observed outcome into a 64-dimensional representation for each context
event. Energy features comprise $L$ sine and cosine pairs
$\gamma_\ell(e)=[\sin(2^\ell\pi e),\cos(2^\ell\pi e)]$ for $\ell=0,\ldots,L-1$.
Table~\ref{tab:architecture-details} summarizes the network dimensions.
All energies used in the physical kernels below are in keV.

\begin{table}[ht]
\centering
\caption{Density-guided CNP architecture. The decoder receives the aggregated
context, raw coordinates, gated Fourier features, and density contrast.}
\label{tab:architecture-details}
\begin{tabular}{ll}
\toprule
Component & Configuration\\
\midrule
Encoder hidden widths & 128, 128\\
Context representation & 64 dimensions per event\\
Attention query and key & Raw $(e,T)$, projected from 2 to 128 dimensions\\
Attention value and output & $64\to128$ and $128\to64$ projections\\
Density maps & Two separate $2\to16\to1$ MLPs with GELU\\
Decoder input & 87 dimensions\\
Decoder hidden widths & 128, 128, 128\\
Decoder output & Logit location $\mu$ and raw scale $\rho$\\
Dropout probability & 0.2\\
Total parameters & 130,693\\
\bottomrule
\end{tabular}
\end{table}

\subsection{Density-guided aggregation and decoding}
The density buffer contains the energies of the matching nominal training
subset, including events excluded by the training sampler. It contains no final
context or reference events. Define the unnormalized kernel sums and the inputs
to the attention maps by
\begin{equation}
A_\sigma(E)=\sum_{p\in\mathcal P}
 \exp\!\left[-\frac{(E-E_p)^2}{2\sigma^2}\right],
 \qquad
 z(E)=\bigl[\log(A_{\sigma_l}(E)+\varepsilon_d),
             \log(A_{\sigma_g}(E)+\varepsilon_d)\bigr].
\end{equation}
Here $\sigma_l$ and $\sigma_g$ are the local and broad density widths, and
$\varepsilon_d$ stabilizes the logarithms and density ratio. Their values are
listed in Table~\ref{tab:model-constants}.
The two independent MLPs $g_h$ and $g_\tau$ have biases and no dropout. They set
the bandwidth and temperature within fixed bounds,
\begin{align}
 h(E)&=h_{\min}+(h_{\max}-h_{\min})\operatorname{sigmoid}(g_h(z(E))),\\
 \tau(E)&=\tau_{\min}+(\tau_{\max}-\tau_{\min})
                     \operatorname{sigmoid}(g_\tau(z(E))).
\end{align}
Because the sums are not
normalized by training-pool size, their log inputs depend on both spectral
shape and sample count.

Attention uses one head of dimension $d_{\mathrm{att}}$. Bias-free projections produce a query
$q$ and keys $k_i$ from raw coordinates, and values $v_i$ from the learned
context representations. Before dropout, aggregation is
\begin{equation}
 a_i(E)=\frac{q^\top k_i}{\sqrt{d_{\mathrm{att}}}\,\tau(E)}
       -\frac{(E-E_i)^2}{2h(E)^2},
 \qquad
 \omega_i(E)=\operatorname{softmax}_i(a(E)),
 \qquad
 r(E)=W_o\sum_i\omega_i(E)v_i .
\end{equation}
We set $d_{\mathrm{att}}=128$.
The implementation applies inverted dropout to the weights without
renormalizing them, followed by dropout after the output projection $W_o$.
Thus the weights sum to one before, but not necessarily after, dropout.

The decoder concatenates $r(E)$, $(e,T)$, the 20 gated Fourier features, and
the scalar density contrast $R(E)$. The main text defines the contrast and
frequency gate. The learned background cutoff is
$\kappa=\kappa_{\min}+(\kappa_{\max}-\kappa_{\min})
\operatorname{sigmoid}(\theta_\kappa)$. The bounds are one and five, and a
zero-initialized $\theta_\kappa$ gives an initial cutoff of three.
The contrast threshold, gate slopes,
frequency cap, density widths, and attention bounds remain fixed during
training. Density guidance encourages an appropriate allocation of local
flexibility.

\begin{table}[ht]
\centering
\caption{Fixed model constants. Symbols separate the estimator's structure
from the numerical configuration used in these experiments.}
\label{tab:model-constants}
\begin{tabular}{lll}
\toprule
Quantity & Symbol & Setting\\
\midrule
Energy origin and scale & $E_{\min},\Delta E$ & 500 keV, 2500 keV\\
Density widths & $\sigma_l,\sigma_g$ & 1 keV, 50 keV\\
Density stabilizer & $\varepsilon_d$ & $10^{-5}$\\
Attention head dimension & $d_{\mathrm{att}}$ & 128\\
Bandwidth bounds & $h_{\min},h_{\max}$ & 5 keV, 200 keV\\
Temperature bounds & $\tau_{\min},\tau_{\max}$ & 1, 10\\
Fourier levels & $L$ & 10\\
Density-contrast threshold & $R_0$ & 3\\
Contrast and frequency slopes & $\alpha_R,\alpha_F$ & 10, 5\\
Background-cutoff bounds & $\kappa_{\min},\kappa_{\max}$ & 1, 5\\
Probability clipping tolerance & $\varepsilon_p$ & $10^{-6}$\\
\bottomrule
\end{tabular}
\end{table}

\subsection{Training and prediction}
The decoder uses $\sigma_j=\operatorname{softplus}(\rho_j)$ and four independent
standard-normal draws per target to form
\begin{equation}
 p_{jm}=\operatorname{clip}_{[\varepsilon_p,\,1-\varepsilon_p]}
 \left[\operatorname{sigmoid}(\mu_j+\sigma_j\epsilon_{jm})\right],
 \qquad \epsilon_{jm}\sim\mathcal N(0,1).
\end{equation}
Here $\varepsilon_p=10^{-6}$ keeps probabilities away from zero and one.
It is distinct from the density stabilizer $\varepsilon_d$ and the random
draw $\epsilon_{jm}$.
Training minimizes the mean binary cross-entropy over these four draws, target
events, and batch trials.  At evaluation, the
reported prediction instead averages 50 dropout-active calls returning
$\operatorname{sigmoid}(\mu_j)$. The evaluation does not sample the
logistic-normal output scale.

Each run uses 3,000 Adam steps with learning rate $10^{-3}$, batch size 16,
and gradient-norm clipping at 1. Trial sizes range from 640 to 1,024 events,
with 128 to 512 context events and thresholds sampled uniformly on $[0,1]$.
The event sampler selects uniformly among 10-keV energy bins containing at
least four training events, then samples events with replacement. No mixup
is applied. We evaluate the final-step checkpoint and do not retune settings
for each training budget.

The base architecture was selected before the controlled budget campaign,
using exploratory models and a development comparison. The 1-keV and 50-keV
density widths reflect detector-scale structure and a broader local
background, respectively. The contrast threshold of 3 was chosen manually
after inspecting a spectral-density diagnostic. The broader kernel can blend neighboring lines.

\subsection{Comparators and mechanism controls}
CNP uses mean aggregation, while Attentive CNP uses query-dependent aggregation.
The attentive baselines are deterministic and omit the latent variable of the
original attentive neural process \citep{kim2019anp}.
The PE comparator adds Fourier features, but its queries and keys use encoded
coordinates and its decoder wiring differs from ours. The main baseline
comparison therefore tests complete architectures rather than isolating
density guidance alone.

The mechanism controls retain the full model's wiring and parameter allocation.
The global-gate control replaces the density-dependent cutoff with one learned
scalar. The global-attention control feeds constant inputs to the two
attention maps, retaining learned global bandwidth and temperature as well as
query-dependent context weighting. A combined control applies both changes,
and the density-free control additionally sets the direct decoder input $R$
to zero. All variants allocate 130,693 parameters, although constant attention
inputs make 64 weights inactive and removing direct $R$ makes another 128
inactive. The restored global cutoffs lie between 3.88 and 4.10.

The mechanism campaign uses three initialization seeds and ten paired contexts
for each variant at the original 5k subset.

\section{Data protocol and supplementary results}
\label{app:evidence}

\subsection{Data allocation and preprocessing}
Table~\ref{tab:data-ledger} separates classifier training, efficiency-model
training, context observations, and reference evaluation. Counts denote unique
events within a pool, not repeated optimizer draws. The efficiency-training
subsets reuse classifier-training events, whereas the final context reservoir
and reference are disjoint from training and from one another. The classifier
is fixed for every efficiency-model comparison.
The source pool contains events from the 16 files \texttt{MJD\_Train\_0.hdf5}
through \texttt{MJD\_Train\_15.hdf5} with $500\leq E<3000$ keV.

\begin{table}[ht]
\centering
\caption{Event budgets and their roles. The 2k, 5k, and 10k labels refer only
to efficiency-model training, conditional on the fixed classifier.}
\label{tab:data-ledger}
\begin{tabular}{lr}
\toprule
Role & Events\\
\midrule
Eligible classifier source & 377,330\\
Classifier training, 5.0\% of source & 18,866\\
Efficiency training, nominal budgets & 2,000 / 5,000 / 10,000\\
Efficiency training, sampler-eligible original subsets & 1,895 / 4,984 / 9,980\\
Threshold calibration & 2,000\\
Development context pool & 3,000\\
Development targets & 2,074\\
Final context reservoir & 20,000\\
Context per final evaluation & 500\\
Shared final reference & 114,400\\
\bottomrule
\end{tabular}
\end{table}

The classifier waveform pipeline subtracts the baseline estimated from the
first 500 samples, normalizes by the positive maximum, and crops around the
first 90\% rise with 200 preceding and 2,000 following samples, using zero
padding where needed. Classifier training uses 50 epochs of binary
cross-entropy with logits and class- and energy-balanced replacement sampling.

The retained classifier is the epoch-50 checkpoint.

The original efficiency-training budgets are nested prefixes of one frozen
ordering. Two additional outcome-blind orderings, with seeds 20260910 and
20260911, provide alternative 5k subsets with 4,981 and 4,980 sampler-eligible
events. These subsets overlap because they come from the same finite pool.
Each neural configuration uses initialization seeds 0, 1, and 2 and the same ten
500-event final contexts. The contexts also overlap.  Settings were frozen
before this repeated-split evaluation.

\subsection{Classical baseline settings}
The context-only kernel estimator predicts the Gaussian-weighted fraction of
passing context events. Its bandwidth selection uses five-fold
cross-validation on development contexts and candidates of 2, 5, 10, 20, 50,
and 100 keV. The pooled-kernel control additionally uses the full 18,866-event
training pool, including events excluded by the neural training sampler.
It therefore has more labeled observations than the context-only estimators.

The Bernoulli GP uses a logistic likelihood and Laplace approximation.
Development Brier scores select between constant-amplitude RBF and
Mat\'ern-$3/2$ kernels, selecting the latter. The completed campaign comprises
80 development fits and 40 final fits across context budgets.

\subsection{Reference bands and bin support}
For each supported 5-keV reference bin $b$, let $n_b$ be its event count,
$\overline X_b$ its observed passing fraction, and $\overline p_b$ the model
prediction averaged at those events' energies. We use the half-width of the
$z=1$ Wilson interval as the common reference scale,
\begin{equation}
 s_b=\frac{\sqrt{\overline X_b(1-\overline X_b)/n_b+1/(4n_b^2)}}{1+1/n_b},
 \qquad
C_k=\frac{100}{|\mathcal B|}
 \sum_{b\in\mathcal B}
 \mathbf 1\!\left[|\overline p_b-\overline X_b|\leq k s_b\right].
\end{equation}
The agreement band is centered at the observed fraction, not the Wilson
interval center.  Every method uses the same reference scale and support rule.

Bins require at least four reference events. Overall coverage uses 442 of
500 bins, excluding 58 bins containing 101 events. The four peak cores each
contain two supported bins. Their centers are 1587.5 and 1592.5 keV for
$^{208}$Tl DEP, 1617.5 and 1622.5 keV for $^{212}$Bi FEP, 2102.5 and 2107.5 keV
for $^{208}$Tl SEP, and 2612.5 and 2617.5 keV for $^{208}$Tl FEP. Thus each
individual peak's coverage in one run can only be 0\%, 50\%, or 100\%, while
averages across runs need not be. Peaks averages the four features equally.
Continuum averages coverage in the 1700 to 2000 and 2200 to 2400 keV windows
equally, rather than weighting their 60 and 40 bins together.

\subsection{Threshold and training-subset sensitivity}
Table~\ref{tab:coverage-budget-sensitivity} retains $C_1$ and $C_3$ alongside
the main tables' $C_2$. The failure at 2k and the tradeoff between peaks and continuum
remain visible across tolerances. At 5k and 10k, the density-guided model
combines stronger peak reconstruction than CNP and Attentive CNP with better
continuum agreement than unrestricted PE. Increasing the training budget
does not improve every endpoint monotonically.

\begin{table}[ht]
\centering
\caption{Reference-band coverage sensitivity (\%). Within each cell,
top/middle/bottom entries give $C_1/C_2/C_3$. All configurations use 500 context
events, three initialization seeds, and ten paired contexts.}
\label{tab:coverage-budget-sensitivity}
\begin{tabular}{lrrrrrr}
\toprule
& \multicolumn{2}{c}{2k training}
& \multicolumn{2}{c}{5k training}
& \multicolumn{2}{c}{10k training}\\
\cmidrule(lr){2-3}\cmidrule(lr){4-5}\cmidrule(lr){6-7}
Model & Peaks & Continuum & Peaks & Continuum & Peaks & Continuum\\
\midrule
CNP & \shortstack{17\\27\\30} & \shortstack{49\\82\\96} & \shortstack{11\\23\\28} & \shortstack{40\\73\\92} & \shortstack{12\\22\\25} & \shortstack{42\\77\\93} \\[2pt]
Attentive CNP & \shortstack{12\\12\\21} & \shortstack{45\\78\\95} & \shortstack{8\\18\\20} & \shortstack{35\\66\\86} & \shortstack{10\\26\\33} & \shortstack{40\\74\\92} \\[2pt]
Attentive CNP + PE & \shortstack{9\\22\\28} & \shortstack{13\\26\\37} & \shortstack{23\\41\\56} & \shortstack{17\\37\\57} & \shortstack{26\\51\\71} & \shortstack{25\\49\\70} \\[2pt]
\textbf{Ours} & \shortstack{5\\13\\30} & \shortstack{20\\42\\56} & \shortstack{20\\45\\62} & \shortstack{51\\89\\99} & \shortstack{17\\42\\70} & \shortstack{51\\88\\99} \\[2pt]
\bottomrule

\end{tabular}
\end{table}

Across the three 5k training subsets, our Peaks $C_2$ is 45.0\%, 32.9\%, and
50.0\%, while Continuum $C_2$ is 88.8\%, 80.4\%, and 84.8\%. All three
subsets outperform the 10k CNP and Attentive CNP on both summaries at each
of $k=1,2,3$. They do not outperform 10k PE at peaks.

\subsection{Supporting mechanism evidence and scope}
For the full model versus the global decoder gate, the paired Peaks $C_2$
gains are $+12.5\%$, $+25.0\%$, and $+35.0\%$ across initialization seeds.
The corresponding Continuum gains are $+4.5\%$, $+8.7\%$, and $+8.0\%$. A continuous
error measure supports the same direction. The equal-feature mean of
event-weighted peak-core absolute error is 5.12\% versus 13.24\%, and
continuum error is 3.96\% versus 4.97\%.

The adaptive gate does not improve every feature. DEP and bismuth FEP improve,
SEP $C_2$ is unchanged, and thallium FEP decreases from 16.7\% to 13.3\%.
Replacing adaptive attention with learned global settings leaves the primary
summaries nearly unchanged. Under global gating and attention, retaining the
direct density input improves Peaks by 2.9\% but lowers Continuum by
5.2\%. The evidence therefore supports adaptive decoder gating most
clearly, rather than independent benefits from every density pathway.

Outside the primary windows, the sparse tail has only 150 reference events
and four passes, and no sampler-eligible training events. Full-model tail
$C_2$ is 53.1\%, versus 70.3\% for the global gate. Equal weighting of the
tail and the two continuum windows gives 76.9\% versus 77.9\%.

\end{document}